\documentclass[num-refs]{nbdt-article}

\newcommand{\figref}[1]{Fig.~\ref{#1}}
\newcommand{\tabref}[1]{Tab.~\ref{#1}}
\newcommand{\secref}[1]{Sec.~\ref{#1}}

\newcommand{\rtwo}{\emph{$R^2\,$}}

\newcommand{\myapprox}{{\raise.17ex\hbox{$\scriptstyle\sim$}}}
\newcommand{\xhdr}[1]{\vspace{0pt}\noindent\textbf{#1}\xspace}

\makeatletter
\newcommand\footnoteref[1]{\protected@xdef\@thefnmark{\ref{#1}}\@footnotemark}
\makeatother

\belowdisplayskip \abovedisplayskip

\usepackage{times}
\usepackage{epsfig}
\usepackage{graphicx}
\usepackage{amsmath}
\usepackage{amssymb}
\usepackage{setspace}
\usepackage{pythonhighlight}
\usepackage{etoolbox}

\usepackage{inconsolata}

\usepackage{multirow}
\usepackage{rotating}
\usepackage{booktabs}
\usepackage{arydshln}

\usepackage{epsfig}
\usepackage{graphicx}
\usepackage{wrapfig}
\usepackage[belowskip=2pt,labelfont=bf, font=small,labelsep=period]{caption}
\usepackage{array}

\usepackage[belowskip=0pt, font=small]{subcaption}
\usepackage[olditem,oldenum]{paralist}

\usepackage{mysymbols}

\usepackage{url}
\usepackage{xspace}
\usepackage{comment}
\usepackage{color}
\usepackage{xcolor}
\usepackage{framed}
\usepackage{fancybox}
\usepackage{caption}

\usepackage{footnote}
\makesavenoteenv{tabular}

\definecolor{citecolor}{HTML}{2779af}
\definecolor{linkcolor}{HTML}{c0392b}
\usepackage[pagebackref=false,breaklinks=true,colorlinks,bookmarks=false,citecolor=citecolor,linkcolor=linkcolor]{hyperref}

\papertype{Original Article}
\paperfield{Neural data science / analysis}

\title{Representation learning for neural population activity with Neural Data Transformers}

\author[1]{Joel Ye}
\author[2]{Chethan Pandarinath}

\affil[1]{School of Interactive Computing, Georgia Institute of Technology, Atlanta, GA, USA}
\affil[2]{Wallace H. Coulter Department of Biomedical Engineering and Department of Neurosurgery, Emory University and Georgia Institute of Technology, Atlanta, GA, USA}

\corremail{joelye9@gmail.com}

\fundinginfo{This work was supported by the Emory Neuromodulation and Technology Innovation Center (ENTICe), NSF NCS 1835364, DARPA
PA-18-02-04-INI-FP-021, NIH Eunice Kennedy Shriver NICHD K12HD073945, the Alfred P. Sloan Foundation, and
the Simons Foundation as part of the Simons-Emory International Consortium on Motor Control.}

\runningauthor{Joel Ye and Chethan Pandarinath}

\begin{document}

\maketitle

\begin{abstract}
Neural population activity is theorized to reflect an underlying dynamical structure. This structure can be accurately captured using state space models with explicit dynamics, such as those based on recurrent neural networks (RNNs). However, using recurrence to  explicitly model dynamics necessitates sequential processing of data, slowing real-time applications such as brain-computer interfaces. Here we introduce the Neural Data Transformer (NDT), a non-recurrent alternative. We test the NDT's ability to capture autonomous dynamical systems by applying it to synthetic datasets with known dynamics and data from monkey motor cortex during a reaching task well-modeled by RNNs. The NDT models these datasets as well as state-of-the-art recurrent models.
Further, its non-recurrence enables 3.9ms inference, well within the loop time of real-time applications and more than 6 times faster than recurrent baselines on the monkey reaching dataset. 
These results suggest that an explicit dynamics model is not necessary to model autonomous neural population dynamics.

Code:~\href{https://github.com/snel-repo/neural-data-transformers}{github.com/snel-repo/neural-data-transformers}.

\end{abstract}

\section{Introduction}

\begin{figure*}
    \centering
    \includegraphics[width=1.0\textwidth]{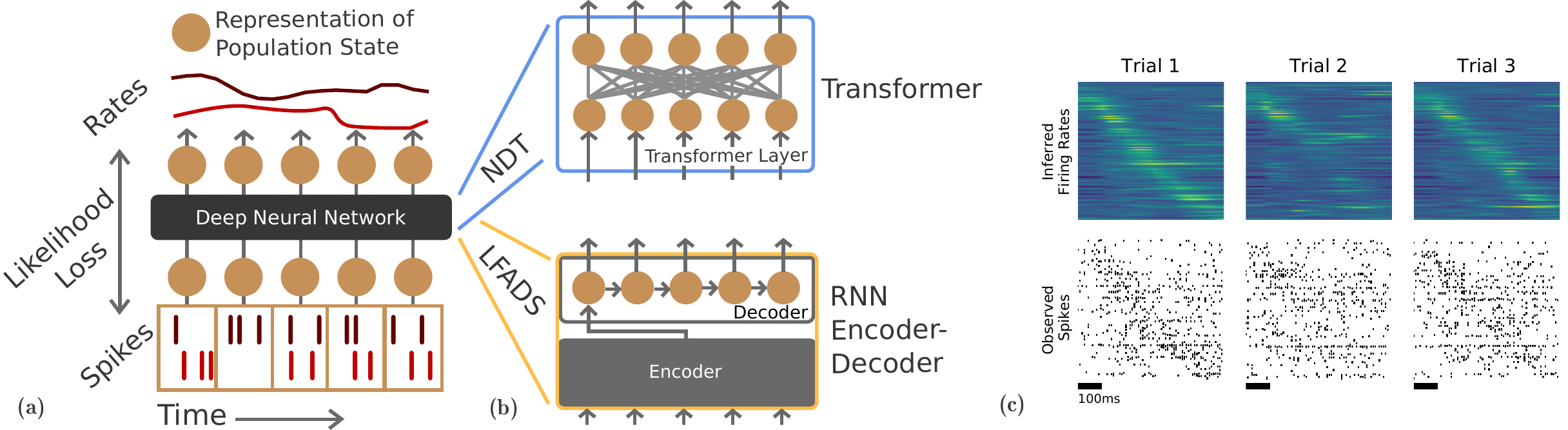}
    \caption{
        \xhdr{Sequential vs. parallel models.} (a) Unsupervised models of sequential spiking activity take in binned spikes (with 2 channels in this schematic) and output inferred rates. A likelihood loss trains the network to output the most likely rates. (b) A Transformer architecture (top) performs parallel modeling, contrasting with RNNs (bottom) and methods like GPFA which use sequential processing. (c) Spike input and inferred rate examples for the NDT applied to the reaching dataset. Rows are sorted by the time of each channel's maximum rate in the first trial, so as to demonstrate correspondence between observed activity and rates.
    }
    \label{fig:schematic}
\end{figure*}

Neural populations are theorized to have an underlying dynamical structure which drives the evolution of population activity over time~\citep{shenoy2013cortical,pandarinath2018jneuro,vyas2020arn}. This structure can be explicitly modeled using linear~\citep{macke2011empirical, kao2015single,gao2016linear} or switching linear dynamical systems~\citep{petreska2011dynamical,linderman2017bayesian}, or nonlinear dynamical systems such as recurrent neural networks (RNNs)~\citep{pandarinath2018inferring,she2020rgp,perich2020rnn}. 
In contrast to traditional analyses that average activity across repeated trials of the same behavior, 
these models have helped relate
neural population activity to behavior in individual trials. In particular, an RNN-based method called latent factor analysis via dynamical systems (LFADS) has been shown to model single trial variability in neural spiking activity far better than traditional baselines like spike smoothing or GPFA~\citep{pandarinath2018inferring,keshtkaran2021autolfads}. This precise modeling enables accurate prediction of subjects’ behaviors on a moment-by-moment basis and millisecond timescale.

RNNs have also been used to model language, and have been analogously shown to capture linguistic structure in input sentences~\citep{tenney2019classical}. However, with the advent of massive language datasets and their costly training implications, the language modeling community has shifted away from recurrent networks and towards the Transformer architecture~\citep{neurips2017vaswani}. A Transformer receives a sequence of word tokens, or inputs, and processes each individual token in parallel. For example, a Transformer can classify the parts of speech of every word in a sentence simultaneously, whereas an RNN must process earlier words before later ones. A Transformer’s parallelism enables it to be trained and operated on sequential data faster than an RNN. Though neuroscience datasets may not yet be large enough to realize much training benefit, reduced inference times could already benefit real-time applications where cycle times are critical, such as brain-computer interfaces or closed-loop neural stimulation.

Here we introduce the Neural Data Transformer (NDT), an architecture for modeling neural population spiking activity. The NDT is based on the BERT encoder~\citep{devlin2019bert} with modifications for application to neuroscientific datasets, specifically multi-electrode spiking activity. Modifications are needed as spiking activity has markedly different statistics than both language data and other time series~\citep{wu2020deep, huang2018music} previously modeled by Transformers. Further, neuroscientific datasets are generally much smaller than typical dataset sizes in other machine learning domains, necessitating careful training decisions~\citep{huang2020fixup}\footnote{
Negative results such as ``difficult training`` are under-reported. Our regularization was inspired by discussion in~\href{https://twitter.com/tim_dettmers/status/1247998807494684672}{this Twitter thread}.
}.

We test the NDT on synthetic and real datasets to validate its performance. In our synthetic datasets, we generate firing rates using autonomous dynamical systems and sample spikes from the firing rates. We show the NDT can use the sampled spikes to recover the unobserved rates as well as LFADS. Further, when applied to activity recorded from monkey motor cortex, NDT-inferred firing rates enable prediction of simultaneously measured behavioral variables as well as rates from LFADS. We then demonstrate the NDT’s inference efficiency, showing it performs inference in 3.9ms with minimal dependence on sequence length. On the monkey dataset, this enables inference $6.7\times$ faster than LFADS. We also include an ablative study measuring the contributions of different design choices, and consider the tradeoffs of using an NDT with fewer layers.

Our results provide a proof-of-principle that recurrence is not necessary to accurately infer neural population firing rates on a single-trial basis, and unlocks for neuroscience an alternative modeling paradigm that has greatly advanced other fields using machine learning models.

\section{The NDT Model}

Both the NDT and LFADS transform sequences of binned spiking activity into inferred firing rates (\figref{fig:schematic}a). In real-time applications, the sequence of spiking activity would come from a rolling window of recent activity that ends with the current timestep. Both models assume a Poisson emission model, meaning inferred rates are compared against the observed spiking activity to compute a Poisson likelihood-based training objective (negative log-likelihood, NLL). 

LFADS is a sequential model (\figref{fig:schematic}b, bottom). In the basic architecture optimized to model autonomous dynamics ~\citep{sussillo2016lfads,pandarinath2018inferring}, LFADS first encodes the full input sequence using a bidirectional RNN (not shown), and this encoding is used to set the decoder's initial state. The decoder then evolves its state across timesteps, modeling the dynamics of the neural population. At each timestep, the decoder's state is transformed directly into rate inferences. LFADS' sequential design, i.e., that the decoder must be initialized with an encoding of the full data window, prevents iterative state updates from a stream of individual timesteps in online inference. We discuss challenges to adapting LFADS into an efficient, iterative RNN further in~\secref{sec:efficiency}, but also discuss benefits of the Transformer beyond speed in~\secref{sec:discussion}. Note also that the challenges in adapting LFADS for iterative inference are further compounded for the more powerful, non-autonomous (input-driven) dynamical architecture ~\citep{pandarinath2018inferring,keshtkaran2021autolfads}.

\begin{figure*}
    \centering
        \includegraphics[width=1.0\textwidth]{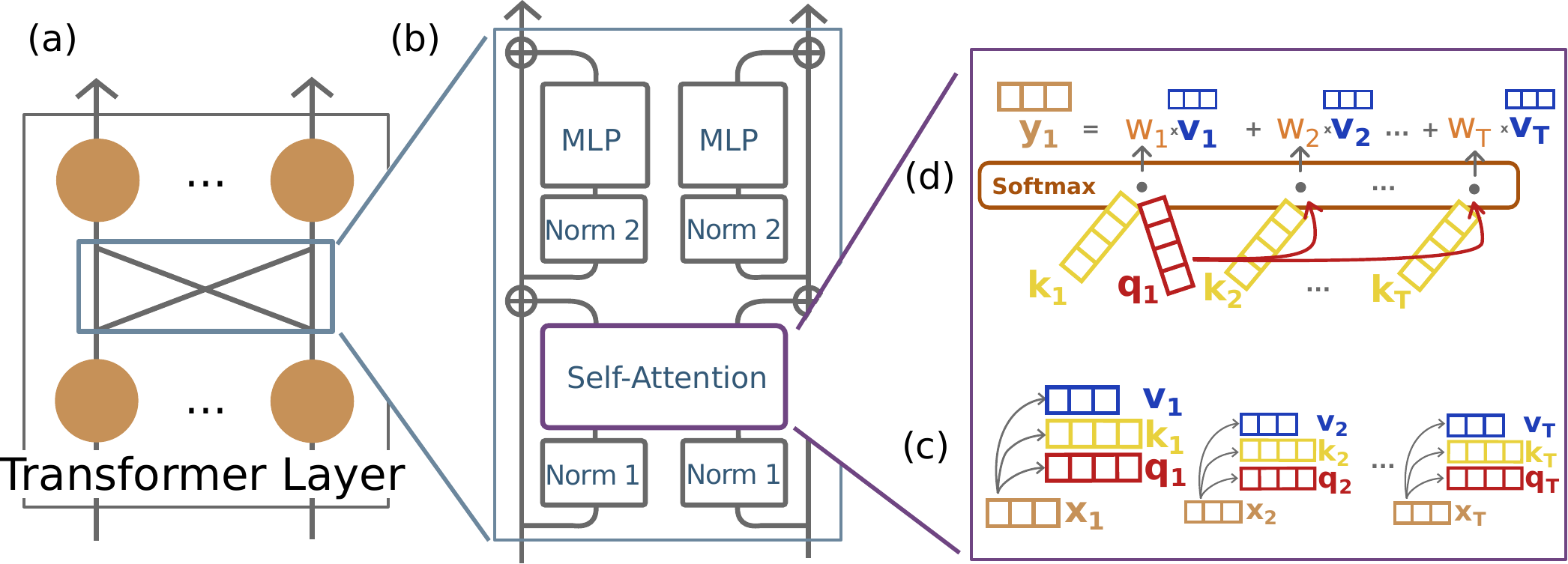}
    \caption{
        \xhdr{Transformer architecture.} (a) A single Transformer layer, as in~\figref{fig:schematic}. The full encoder stacks several of these layers. (b) Inputs to Transformer layers are normalized (``Norm'' blocks), enriched through contextual information (``Self-Attention'' blocks), and passed through a feedforward module (``MLP'' \ie multi-layer perceptron blocks). Blocks with the same label share parameters. The circled plus symbols indicate addition.
        (c) Inputs at each time step are multiplied by three learned weight matrices (not shown) to create three sets of vectors: the queries, keys, and values. 
        (d) Assembling a single timestep's output: Dot products are computed between the timestep's query and every key, yielding similarity scores. Scores are normalized to sum to 1 to form weights. With these weights, a weighted sum of value vectors are computed and returned as the timestep's output.
    }
    \label{fig:details}
\end{figure*}

The Transformer avoids sequential bottlenecks by using a stack of layers that process all inputs together (\figref{fig:schematic}b, top, decomposes one such layer). A Transformer layer comprises several nonlinear sub-layers or blocks (\figref{fig:details}b), in particular a self-attention block in which a new representation of each input is constructed by incorporating relevant information from every other input. We next detail this self-attention block.

The self-attention block is the only one that simultaneously transforms multiple inputs, $x_{[1:T]}$, into multiple outputs, $y_{[1:T]}$. It comprises three different learned weight matrices $W^Q, W^K, W^V$, termed the query, key, and value weights. All inputs are multiplied by these weight matrices to form three sets of intermediate representations, correspondingly termed the queries, keys, and values $q_{[1:T]}, k_{[1:T]}, v_{[1:T]}$, shown in~\figref{fig:details}c. For example, $q_i = W^Qx_i$. For simplicity and to enable stacking of Transformer layers, dimensionality is often fixed throughout the Transformer, \eg given $x_i$ of dimensionality $d$, $W^Q, W^K, W^V$ are $d\times d$ matrices.

An output $y_i$ is formed by taking a weighted sum of the values, with weights $w_i^{[1:T]}$. Each weight $w_i^j$ represents the ``attention'' that step $i$ pays to step $j$ (\figref{fig:details}d), and $w_i^j$ is determined by calculating dot-product similarity between query $i$ and key $j$, and then normalizing similarities over all $j$ with the softmax function. Formally:
\begin{align*}
    y_i &= \sum_{j=1}^T w_i^j v_j \\
    p_i^j &= q_i \cdot k^j \\
    w_i^j &= \frac{\exp{(p_i^j)}}{\sum_{l=1}^T \exp{(p_i^l)}}
\end{align*}
In matrix notation, we can summarize the transformation, from inputs $X$ to outputs $Y$ as: 
\begin{align*}
    Q &=W^QX, K = W^KX, V = W^VX \\
    Y &= \text{softmax}(QK^T)V
\end{align*}

It is possible that inputs may want to attend to each other in multiple complementary manners, which may be difficult to capture with a single set of query, key, and value matrices. Consequently, Transformers are often used with multi-headed attention, where each "head" comprises one query, key, and value matrix set. With $h$ heads, we will get $h$ sets of outputs. To preserve dimensionality, we simply concatenate these $h$ outputs and linearly project them to the input shape. While most language Transformers benefit from $h \geq 8$, we do not find multi-headed attention useful in our experiments, so we keep to small values of $h=1,2$.

The body of the NDT architecture is a Transformer encoder with 6 layers in most of our experiments, as in~\citet{neurips2017vaswani}. We briefly discuss the option to use fewer layers in~\secref{sec:efficiency}. Before entering the encoder, each channel of the observed activity $s_i$ can be optionally projected to an $n$-dimensional embedding, \ie, for activity with $C$ channels, the dimension of each input representation is then $Cn$. To keep dimensionality small, we only consider $n \in \{1, 2\}$ in our experiments. We pass the Transformer encoder outputs through a linear layer and exponentiation (thus treating the linear layer outputs as log-firing rates) before calculating the NLL. Instead of the cross-entropy loss used in language modeling, these log-firing rates are passed into a Poisson likelihood loss. Full notes on the Transformer encoder are provided in~\secref{sec:methods_arch}.

\begin{figure}
    \centering
    \includegraphics[width=0.3\textwidth]{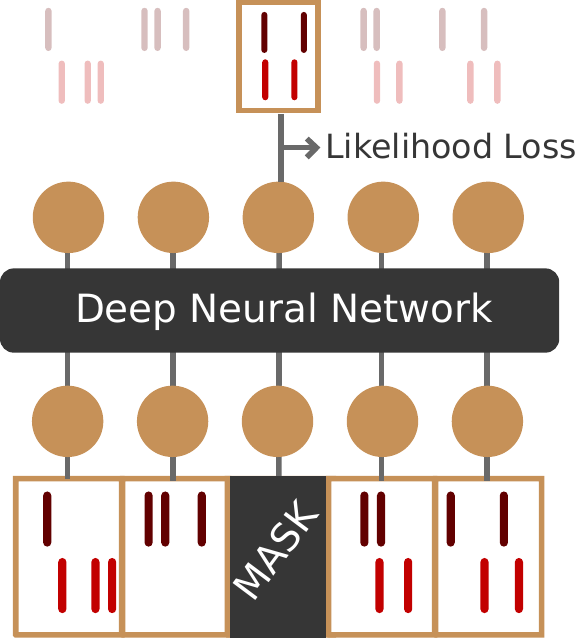}
    \caption{
        \xhdr{Transformer training.} The model is trained with masked modeling~\citep{devlin2019bert,neurips2019keshtkaran}, that is, model outputs are optimized to maximize likelihood of the masked activity given the context provided by unmasked activity.
    }
    \label{fig:training_details}
\end{figure}

To train the model in an unsupervised manner, we adapt the masked modeling methodology used in BERT (\figref{fig:training_details}). In masked modeling, the model is given an input sequence $s_1 \dots s_T$, with a random subset of the $T$ input tokens masked. Subset size is typically a fixed ratio of the full sequence, \eg, 20\% of the inputs. The model is then asked to reproduce the original input for that masked subset. To do so, the model must learn how to leverage the context provided by the unmasked timesteps (\eg, if firing rates in the dataset are temporally smooth, high spike counts in unmasked timesteps may imply high spike counts in masked timesteps). Readers familiar with LFADS might note that masked modeling resembles the coordinated dropout method developed to regularize LFADS models~\citep{neurips2019keshtkaran}, only differing in that coordinated dropout masks individual dimensions (channels) of a given input timestep independently and is not constrained to mask entire input timesteps.

We adjust the training procedure as follows:
\begin{itemize}
    \item In BERT, masked inputs are typically replaced with a special ``[MASK]'' token. Instead of using this special token, which introduces a large distribution shift between training and inference time~\citep{devlin2019bert}, we use a ``zero mask.'' That is, we simply zero out the spike inputs of a masked timestep, which was previously demonstrated to be an effective masking strategy for spiking data~\citep{neurips2019keshtkaran}.
    \item We use intensive regularization to stabilize training, which we find especially important when dataset sizes are smaller. Specifically, in the dropout layers (see \secref{sec:methods_arch} for locations), dropout ratios are swept~$\in[0.2,0.6]$.
\end{itemize}

The importance of the design choices presented here are proven in an ablative study in~\secref{sec:ablative}.

\section{Results}
\label{sec:results}
\begin{figure*}
    \centering
    
    \begin{subfigure}{.52\textwidth}
        \centering
        \includegraphics[width=1.0\textwidth]{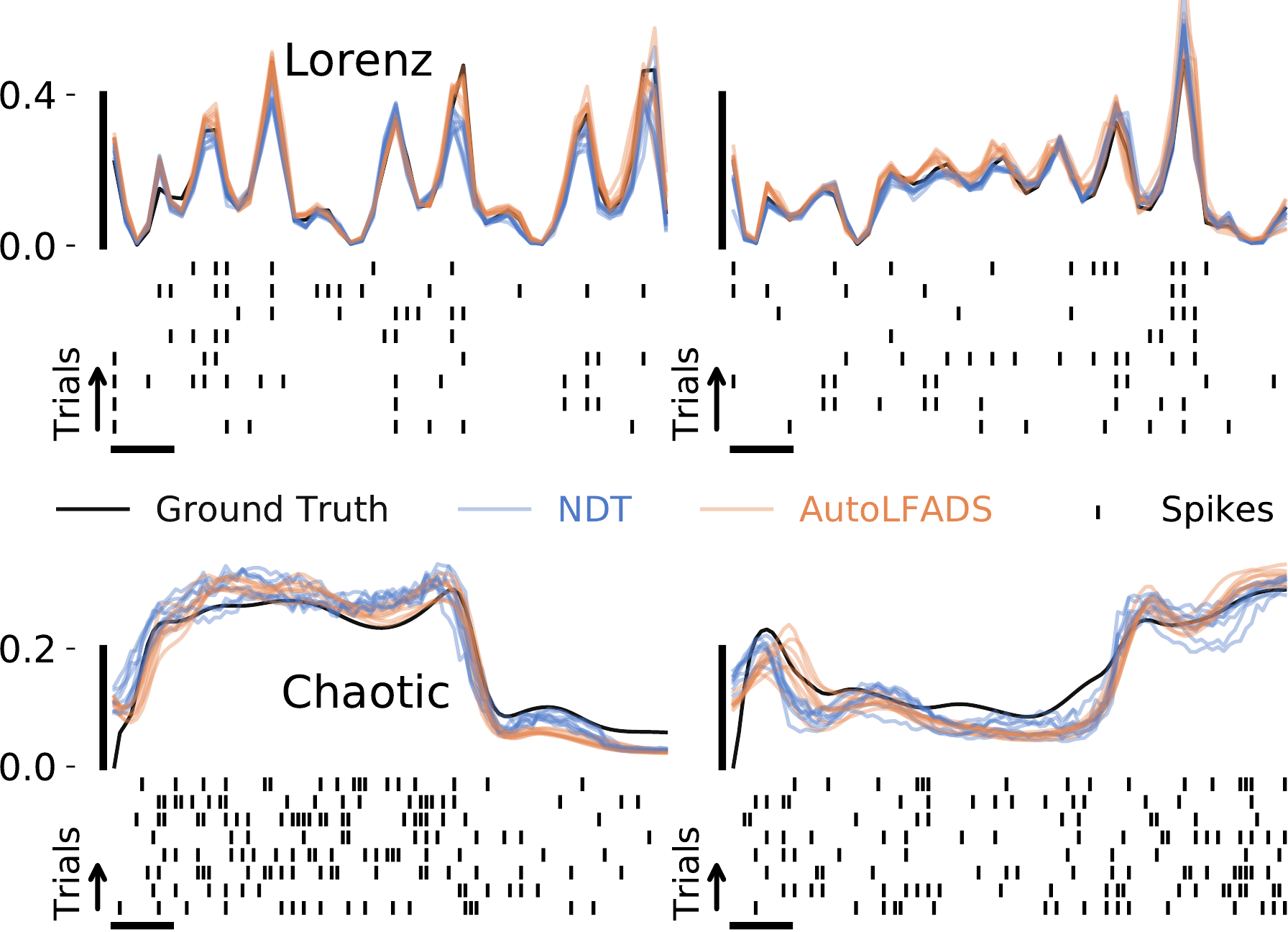}
        \caption{}\label{fig:synth_qual}
    \end{subfigure}
    \hspace{4ex}
    \begin{subfigure}{0.26\textwidth}
        \centering
        \includegraphics[width=1.0\textwidth]{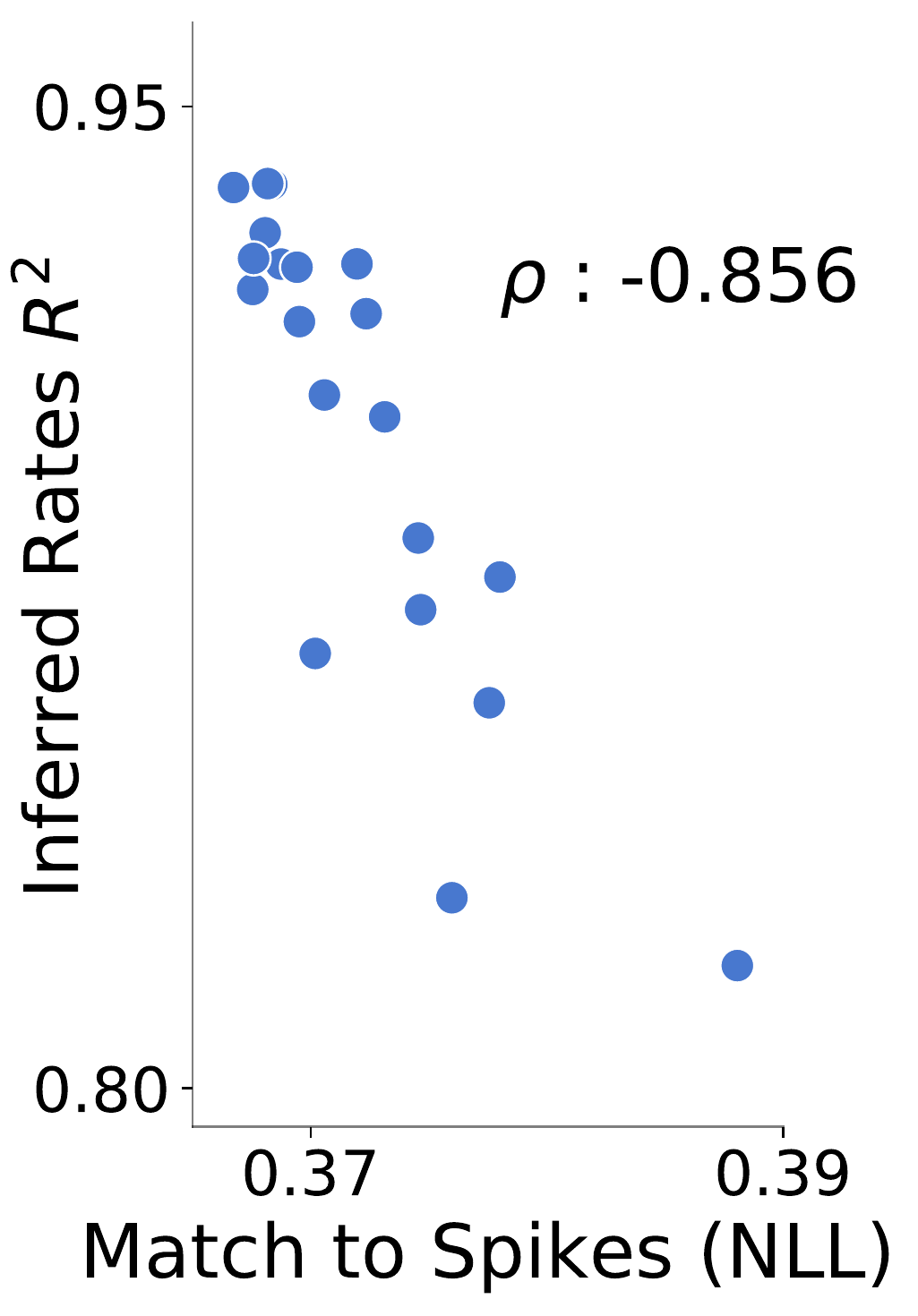}
        \caption{}\label{fig:match_spikes}
    \end{subfigure}
    \caption{
        \xhdr{Modeling synthetic data.} (a) In each quadrant, we plot the ground truth firing rate for a sampled neuron and information from 8 of its trials. We refer to these trials with the same initial conditions (and thus firing rates) as having the same condition and show 2 conditions (columns) for each of the synthetic datasets (rows). For these trials, we show the generated spikes (bottom) and inferred rates from AutoLFADS and NDT (top). Inferred firing rates closely match generating ground truth rates. Vertical bar denotes spikes per bin. Horizontal bar indicates 10\% of the trial length, 5 bins for Lorenz and 10 bins for Chaotic RNN. (b) Across a hyperparameter sweep on the Lorenz dataset, models that achieve better likelihoods yield more accurate inference of the underlying rates. NLL is averaged across bins and channels.
    }
    \label{fig:synth_3}
\end{figure*}

We compare the NDT with LFADS on both synthetic autonomous dynamics and M1 reaching activity, optimizing hyperparameters (HPs, ranges in~\secref{sec:methods_hp}) as follows:
\begin{itemize}
    \item NDT is optimized using grid search. Using early stopping, we select the checkpoint with least validation NLL as measured without masking.
    \item LFADS is optimized using the AutoLFADS framework~\citep{keshtkaran2021autolfads}. LFADS is known to benefit from Population-Based Training (PBT~\citep{jaderberg2017population}) over simple grid search. (We find that NDT performs comparably between grid search and PBT.) AutoLFADS PBT is run with exponentially-smoothed validation NLL as the exploitation metric, and so we select the least smoothed validation NLL checkpoint~\citep{keshtkaran2021autolfads}.
\end{itemize}
Each search has 20 models. We run three searches for each experiment (a total of $3 * 20 = 60$ models are trained) and report the mean and SEM of the metrics achieved by the best model. We select these best models according to likelihood, since likelihood does not require knowledge of the underlying system and is measurable in both synthetic and real-world settings. The NLL reported is averaged over all bin-channel observations. We apply AutoLFADS with fixed settings that were previously shown to work in a variety of applications~\citep{keshtkaran2021autolfads}. Note that our goal is to use AutoLFADS to provide a baseline for comparison, and we do not exhaustively explore its design choices or alternate hyperparameter ranges to achieve a performance ceiling or minimize training/inference times. 

\subsection{The NDT achieves high-fidelity inference on synthetic autonomous dynamical systems}
\label{sec:synthetic-results}

We first evaluate the NDT on two synthetic datasets where observed activity reflects autonomous dynamics: the Lorenz system and the chaotic RNN (details in~\secref{sec:methods_synth}). A trial in the Lorenz system ~\citep{sussillo2014neural,zhao2017variational} is created by simulating a 3D state evolving according to the Lorenz equations, and projecting it to a specified higher dimensionality to form firing rates for a population of synthetic neurons. These rates are sampled according to a Poisson distribution to generate spikes. To create the dataset, we sample several initial states, generate firing rates for each initial state, and sample several trials of spiking activity for each set of firing rates. We denote trials with the same initial state as having the same condition.
Similarly, the chaotic RNN dataset~\citep{sussillo2016lfads} is created by simulating dynamics using a vanilla RNN whose weights are initialized from the normal distribution. This system is motivated by the fact that many neural datasets are well modeled by RNNs (which are themselves nonlinear dynamical systems). The chaotic RNN is more complex than the Lorenz system - as measured by the number of principal components underlying the generating system - and is thus more challenging to model. As with the Lorenz dataset, conditions in the chaotic RNN dataset correspond to distinct initializations of the nonlinear dynamical system.
\begin{table}
  \setlength{\tabcolsep}{4pt}
    \centering
  \resizebox{0.5\linewidth}{!}{
    \begin{tabular} {l c c}
    \toprule
     Dataset & NDT ($R^2$ $\mathrel{\raisebox{0.1ex}{$\uparrow$}}$) & LFADS ($R^2$ $\mathrel{\raisebox{0.1ex}{$\uparrow$}}$) \\ 
      \midrule 
       Lorenz & $0.934$ \scriptsize{$\pm 0.004$} & $0.921$ \scriptsize{$\pm 0.005$} \\
       Chaotic & $0.846$ \scriptsize{$\pm 0.011$} & $0.869$ \scriptsize{$\pm 0.001$} \\
        \bottomrule 
    \end{tabular}
}
    \caption{The NDT and AutoLFADS both infer firing rates that closely align with the generating firing rates on synthetic datasets.}
    \label{tab:synthetic}
\end{table}

The synthetic setting allows us to evaluate inferred firing rates by comparing against the ground truth rates that produced the synthetic spikes. In both datasets, NDT and AutoLFADS inferences closely match the ground truth, though NDT rates appear less smooth (\figref{fig:synth_qual}). We quantify model inference quality by measuring the correspondence between inferred and ground truth firing rates using the coefficient of determination (\rtwo; \tabref{tab:synthetic}). In both datasets, the gap between the two models is small, indicating the NDT can accurately infer firing rates in autonomous dynamical systems. Importantly, we also find that within an HP search, NDT models with high data likelihoods (as computed on the observed spiking activity) tend to match the underlying systems well (as measured by correspondence with ground truth firing rates, \figref{fig:match_spikes}). This match between likelihoods and firing rate inference does not occur in LFADS models that lack coordinated dropout~\citep{neurips2019keshtkaran} and provides a key confirmation that the NDT’s masking strategy works as desired. Verifying that likelihood correlates with recovery of underlying structure in synthetic data provides confidence that likelihood can be used to optimize and choose between NDT models in applications to real-world data.

\subsection{NDT infers motor cortical firing rates in autonomous settings with high fidelity}
To test performance in real-world neural recordings, we apply NDT to the Monkey J Maze dataset~\citep{kaufman2016largest}. These data were previously used to evaluate LFADS and AutoLFADS~\citep{pandarinath2018inferring,neurips2019keshtkaran,keshtkaran2021autolfads} and serve as a benchmark for models of autonomous dynamics. In this dataset, spiking activity from 202 neurons in the primary motor and dorsal premotor cortices was recorded as a monkey performed a delayed reaching task with a variety of straight and curved reaches. 
\newpage
\begin{figure*}
    \centering
    \begin{tabular}[t]{cc}
        \begin{subfigure}{.47\textwidth}
            \centering
            \includegraphics[width=1.0\textwidth]{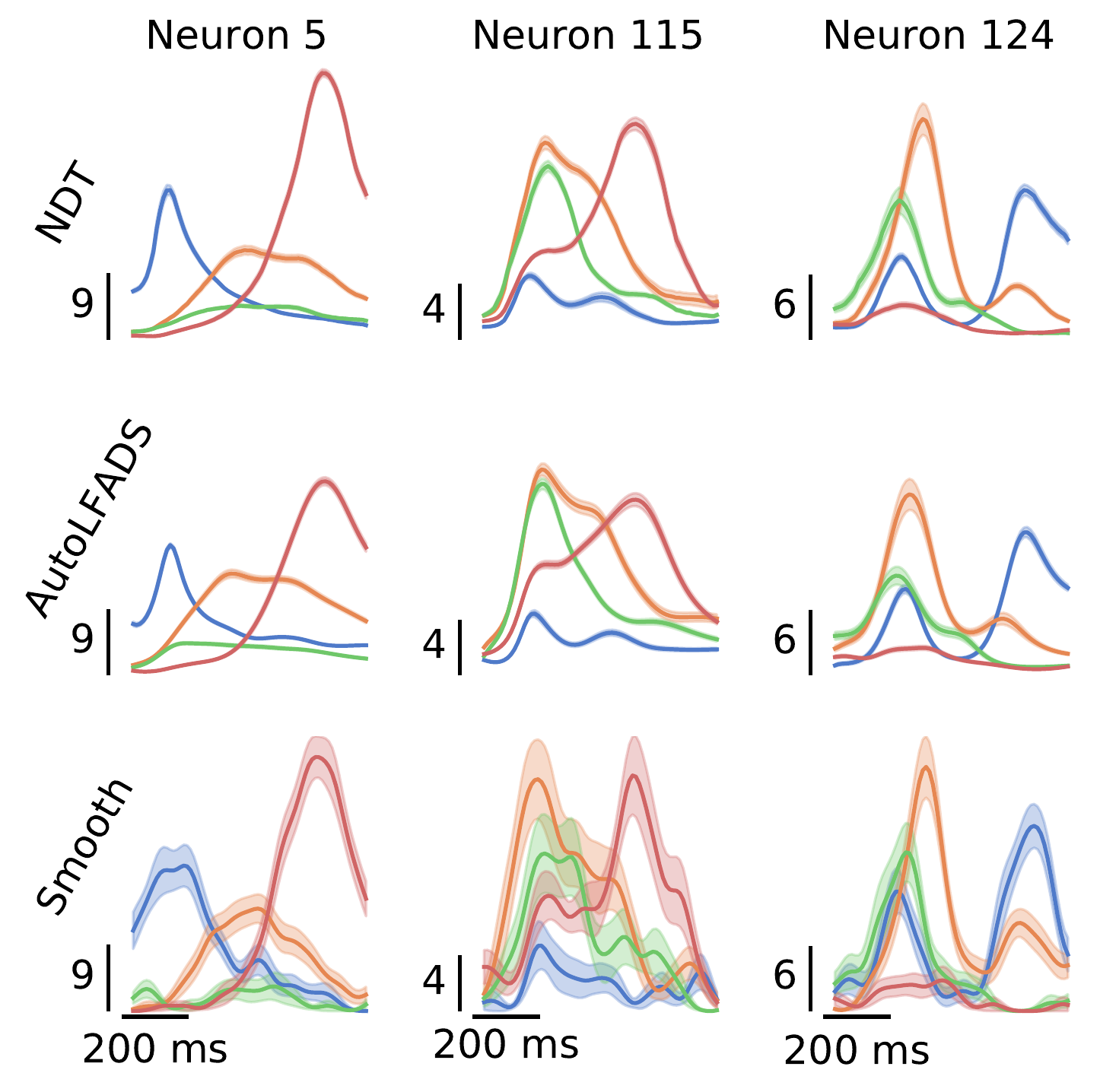}
            \caption{}\label{fig:maze_qual}
        \end{subfigure}
        &
        \begin{subfigure}{0.43\textwidth}
            \centering
            
            \begin{subfigure}{0.48\textwidth}
                \centering
                \includegraphics[width=1.0\textwidth]{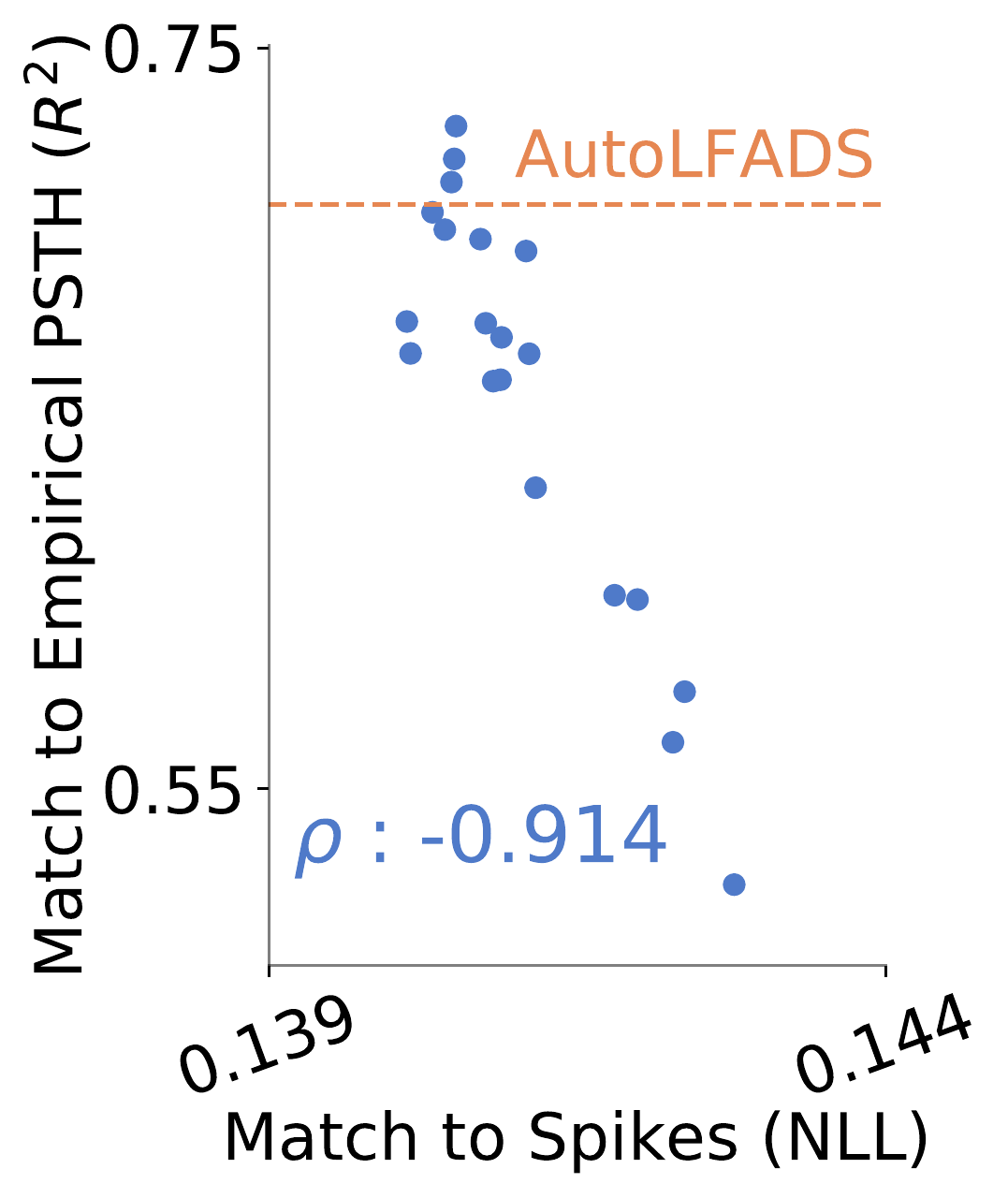}
                \caption{}\label{fig:match_psth}
            \end{subfigure}
            \begin{subfigure}{0.42\textwidth}
                \centering
                \includegraphics[width=1.0\textwidth]{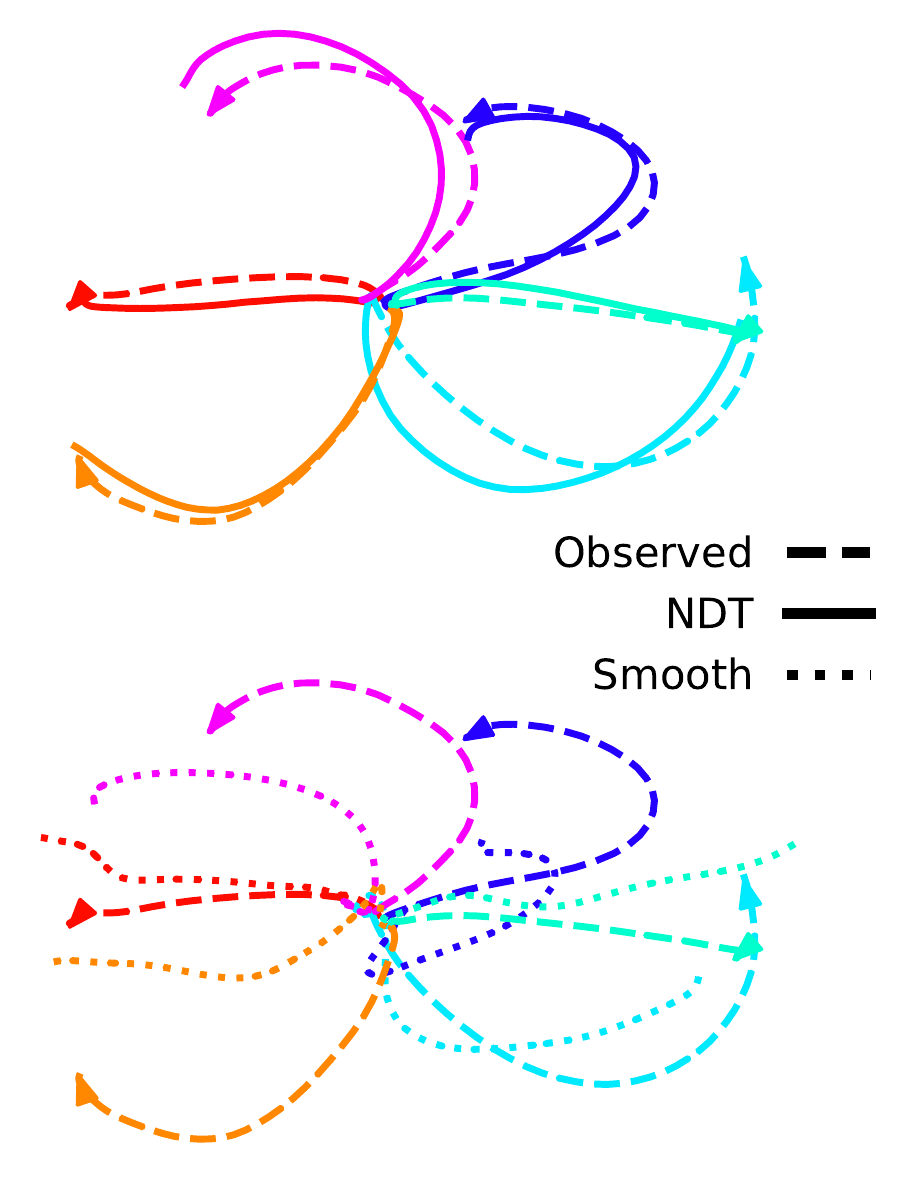}
                \caption{}\label{fig:reaching_traces}
            \end{subfigure} \\
            
            \begin{subfigure}{0.9\textwidth}
                \centering
                \includegraphics[width=1.0\textwidth]{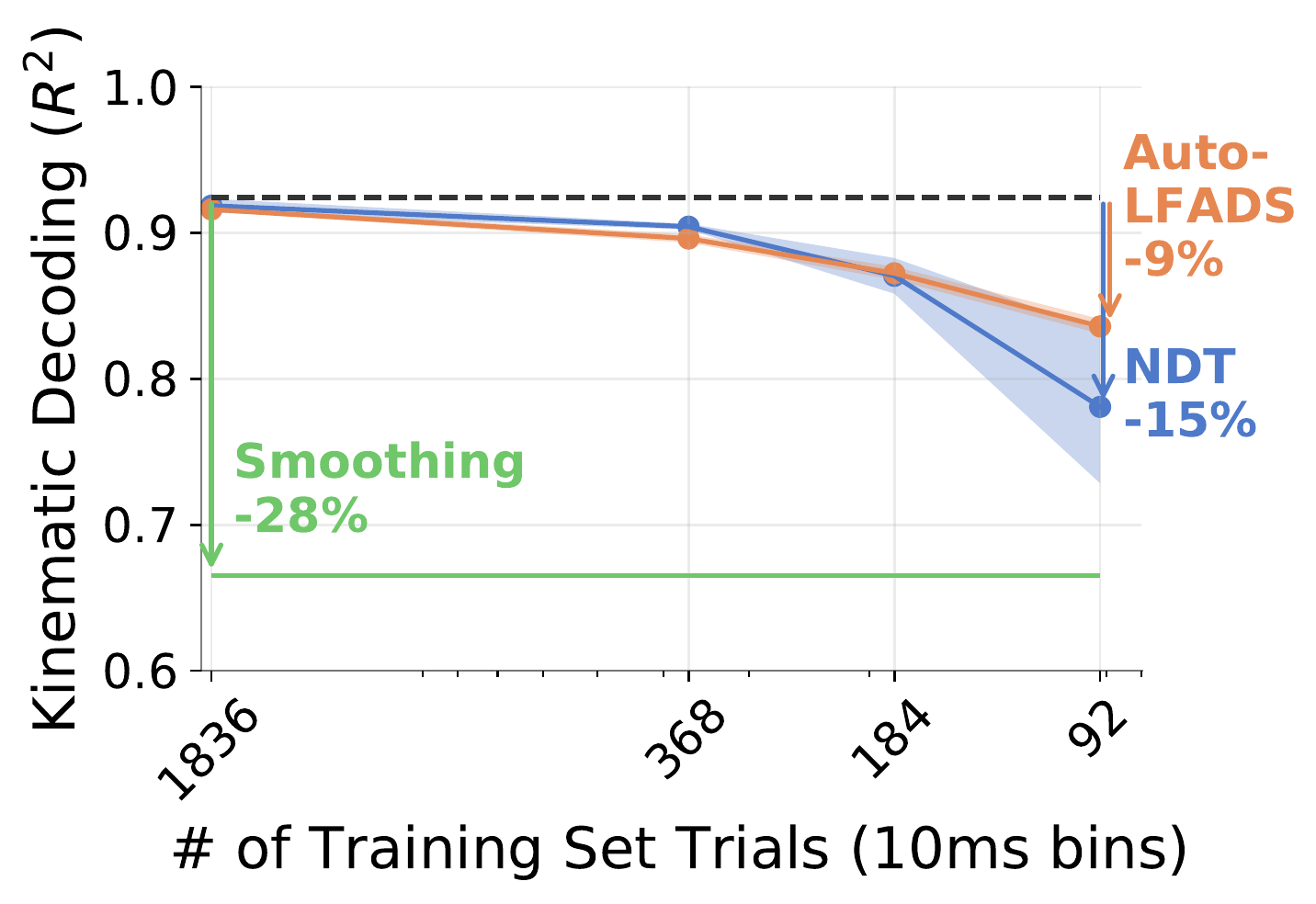}
                \caption{}\label{fig:data_ablate}
            \end{subfigure}
        \end{subfigure}
    \end{tabular}
    \caption{
        \xhdr{NDT inference on monkey reaching data.} (a) Trial-averaged inferred rates and smoothed activity (4 of 108 conditions shown). Vertical bar denotes spikes/sec. Shading indicates SEM. (b) Across a hyperparameter sweep, models that converge to better likelihoods tend to achieve better match to empirical PSTHs. The NLL reported is averaged over the NLL of all bin-channel observations. (c) Example predicted reach trajectories based on NDT-inferred rates (dashed) align more closely with the true trajectories (solid) than predictions based on smoothed spikes (dotted). (d) Performance on kinematic decoding with smaller training sets still greatly improves on spike smoothing down to 92 training trials. Shading indicates SEM across three sweeps.
    }
    \label{fig:m1_3}
\end{figure*}

The reaching dataset consists of 2296 trials across 108 different reach conditions, where a given condition is specified by targets and obstacles present (session 2009-09-18 of \cite{maze_datarelease}). Each trial has a random delay period that separates target presentation from a ``Go'' cue that prompts the monkey to begin its reach, which provides a time period for the monkey to plan before executing the reach. Previous analyses of this paradigm demonstrated that neural activity is well modeled as an autonomous dynamical system, where plan activity serves as an initial state that predicts the activity patterns observed during movement execution~\citep{pandarinath2018inferring,churchland2012neural,shenoy2013cortical}. We train our models on activity during this autonomous period, spanning 250 ms before movement onset to 450 ms after. We perform most experiments by binning the spike sequences at 10ms
; we find similar 
results for bin sizes varying from 2ms to 20ms (results not shown). 

We compute peri-stimulus time histograms (PSTHs) for the models by averaging inferred rates across repeated trials of the same reach condition (\figref{fig:maze_qual}). Both NDT and AutoLFADS exhibit low across-trial variance (as shown by the shaded errorbars), indicating that the models produce consistent inferred rates for different trials of the same condition. We also calculate a spike smoothing baseline by first passing observed spiking activity through a Gaussian kernel with 30 ms standard deviation, and then averaging across trials to form empirical PSTHs (\figref{fig:maze_qual}, bottom). These exhibit larger across-trial variance than the model-inferred firing rates, as spike smoothing produces noisy estimates on single trials~\citep{pandarinath2018inferring}. To quantify the quality of the models’ inferred rates, we measure the correspondence between inferred PSTHs and empirical PSTHs. NDT models with greater likelihoods tend to have better \rtwo (\figref{fig:match_psth}). The highest-performing models perform on par with AutoLFADS.

For motor cortical datasets, another method to evaluate the quality of inferred firing rates is through behavioral decoding, \ie, testing how well simultaneously-recorded behavioral variables can be decoded from the models’ inferred rates. We use optimal linear estimation to map firing rates onto hand velocities (details in~\secref{sec:methods_kin}) and find that NDT enables accurate behavioral decoding that matches AutoLFADS ($0.918$ and $0.915$ \rtwo, respectively). These velocity predictions can be integrated to produce predicted reaching trajectories (\figref{fig:reaching_traces}). The large number of trials (2000) also allows us to evaluate each model’s sensitivity to dataset size by subsampling from the full dataset. NDT comfortably outperforms the spike smoothing baseline, even when scaling to as few as 92 training trials (\figref{fig:data_ablate}). While a 6-layer NDT performs worse than AutoLFADS at 92 trials, we show that a 2-layer NDT closes the gap in~\secref{sec:smaller}.

\subsection{Efficiency Gains from Parallelism}
\label{sec:efficiency}

\begin{figure}
    \centering
    \begin{subfigure}{0.4\textwidth}
    \centering
    \includegraphics[width=0.9\textwidth]{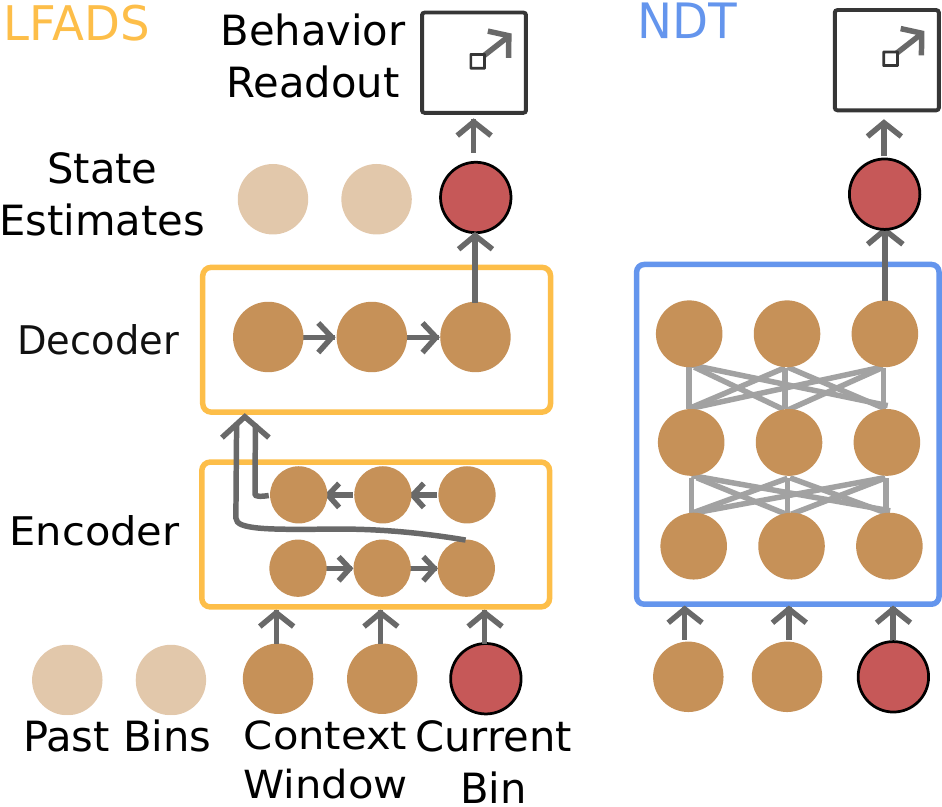}
    \caption{}
    \label{fig:streaming}
    \end{subfigure} %
    \begin{subfigure}{0.45\textwidth}
    \centering
    \includegraphics[width=1.0\textwidth]{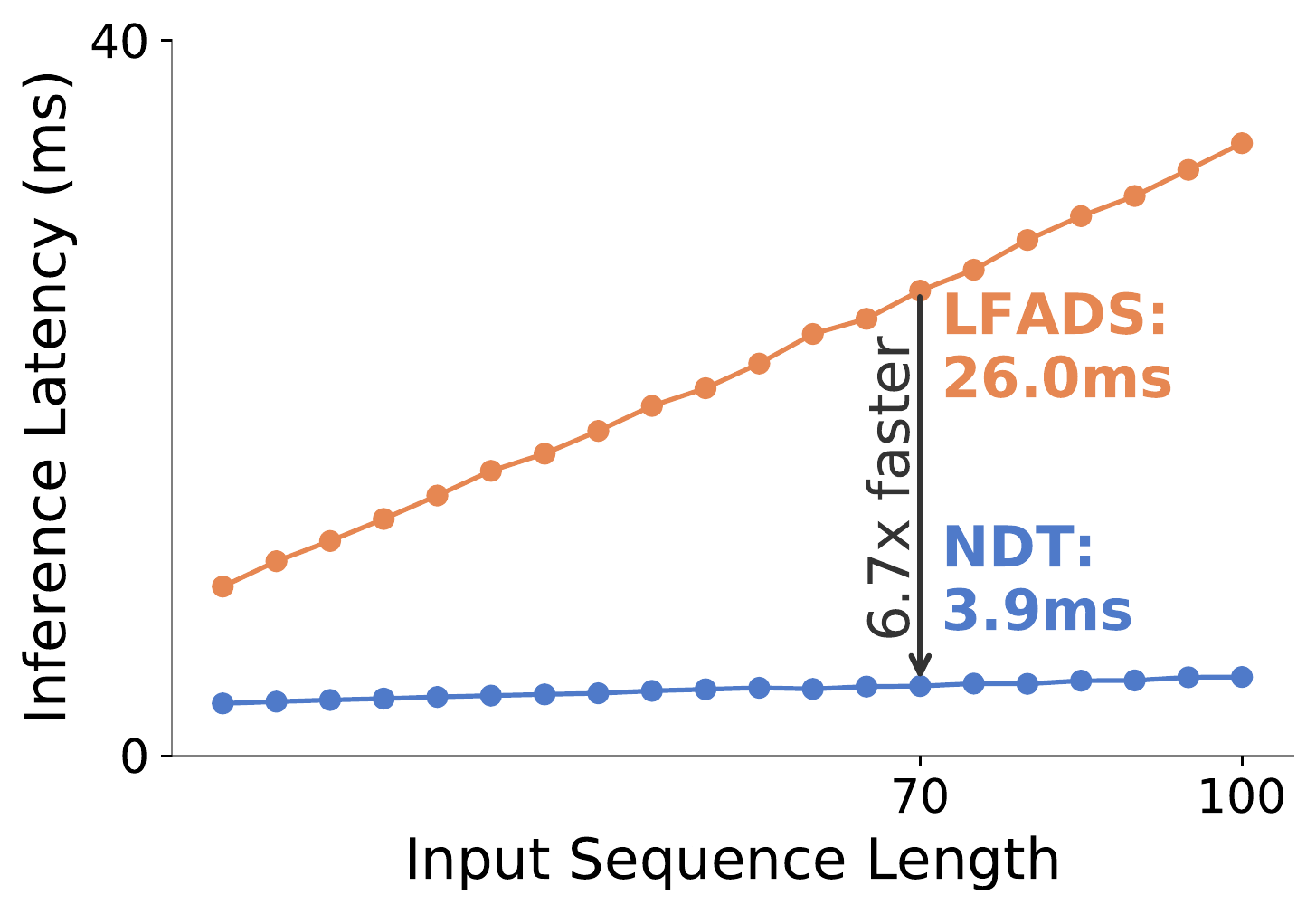}
    \caption{}
    \label{fig:efficiency}
    \end{subfigure}%

    \caption{
    (a) Schematic of LFADS (simplified) and NDT performing online inference.
    (b) Sequential vs. parallel runtime: Recurrent architectures, such as LFADS, have inference times that grow with input sequence length. The NDT has near-constant runtime. In the reaching dataset with input sequences of length 70, the NDT infers $6.7\times$ faster than LFADS.
    }
\label{fig:efficiency_panels}
\end{figure}

\xhdr{Inference Speeds.}
In neural prosthetic applications such as brain-machine interfaces (BMIs) or closed-loop neural stimulation, neural population states are used to guide actions, such as steering a computer cursor or robotic arm, or deciding whether or how to stimulate the brain. In these settings, where neural activity is read-in and acted upon in real-time, even delays on the scale of milliseconds can decrease the quality of closed-loop control \cite{cunningham_closed-loop_2010,gilja_clinical_2015,grosenick_closed-loop_2015,bolus_design_2018,bolus_2020_optimalfeedback}. Thus, inference latencies present a critical barrier to translating computationally-intensive methods for neural state estimation to real-time applications.

We reasoned that the NDT could enable substantial speedups in neural population state inference relative to LFADS, due to the latter architecture's reliance on sequential processing. LFADS and NDT both require a span of observed activity to infer the neural population state at the current time step. To minimize latency in real-time applications, this context would come from a rolling window of observed data (\figref{fig:streaming}). However, despite the overlap in consecutive windows, both models would need to process each window from scratch. LFADS requires a bidirectional reading of the entire input sequence to initialize its dynamics model, which is then run forward for the full input length to infer the state at the current time step.
Note that a new initial condition and subsequent unrolling may not be necessary in autonomous settings (which we use in this work) as activity beyond a certain length into the future should not inform the initial state. However, in real-time, non-autonomous applications, LFADS requires the full context for its controller to make precise interventions in the unrolled dynamics~\cite{pandarinath2018inferring,sussillo2016lfads}.
Similarly, for a multi-layer NDT, most computations would need to be performed anew to take into account the most recent data, and thus we would not expect a large benefit from maintaining state across windows. Consequently, we can compare inference speeds in the offline setting and expect trends to hold in online translation.

\begin{table}
  \setlength{\tabcolsep}{4pt}
    \centering
  \resizebox{0.8\linewidth}{!}{
    \begin{tabular} {l c c c c}
    \toprule
      & Train Time & Inference Time & \# Parameters & Reaching \rtwo \\ 
      \midrule 
       AutoLFADS & 45m & 26ms & 280K & $0.915$ \\
       NDT-6 & 9.4hr & 3.9ms & 1.36M & $0.918$ \\
       NDT-2 & 45m & 2.2ms & 480K & $0.921$ \\
       NDT-1 & 20m & 0.98ms & 270K & $0.886$ \\
    \bottomrule 
    \end{tabular}
}
    \caption{\xhdr{Speed gains from reducing model size.} On one sweep for each variant, we report the training time of the best model, inference time on 70 bins, number of trainable parameters, and kinematic decoding performance.}
    \label{tab:stats}
\end{table}

We measured the inference time for each model as a function of sequence length. Since the NDT models a given input sequence in parallel, we should expect a roughly constant inference speed with respect to input sequence length. The NDT’s non-recurrence enables 3.9ms inference (\figref{fig:efficiency}, with details in~\secref{sec:methods_time}), comfortably within the loop time of many real-time applications. In practice, we find the NDT’s inference times increase slightly with increased bin lengths; in contrast, LFADS inference times increase substantially. In the reaching dataset with sequence length 70, this amounts to a 6.7x speedup. Thus if a recording module reported activity every $x=10$ms, an NDT-based BMI could reflect the latest activity in $10+4=14$ms, whereas an LFADS-based BMI would use $10+26=36$ms. Smaller $x$ yields further benefits. For reference, prior work that achieved high-performing online decoding uses windows with 20 bins of 15ms~\citep{kao2015single}; even with this reduced bin count our method provides a 4x speedup.

For completeness, we note that it may be possible to speed up a recurrent architecture like LFADS if it could maintain state across windows, as is done in traditional iterative state space models such as Kalman filters. The changed model would then only need to integrate one new bin of information, which should reduce the required inference time. However, to our knowledge, such an approach has not yet been demonstrated, suggesting that training such an iterative model is non-trivial.
For example, LFADS' bidirectional encoding, which precludes iterative updates, is a design choice empirically needed for good performance: in unpublished tests, we find that forward-only encoding diminishes performance.

\subsubsection{Smaller NDTs Improve Training Speed and Data Efficiency}
\label{sec:smaller}
The fixed computational complexity of the NDT’s parallel architecture should grant faster training in addition to inference~\citep{neurips2017vaswani}. The 6-layer NDT used in previous experiments, however, does train for significantly longer than our LFADS model~(\tabref{tab:stats}). We note that training times of different models across an HP search can vary widely, \ie, we see NDT 6-layer times between 3 and 18 hours. However, training times can be reduced substantially by simply using a smaller NDT. We find a 1-layer NDT, with around the same number of parameters as our LFADS model, trains under 30 minutes (and infers in under 1ms). Remarkably, this 1-layer NDT achieves $0.89$ \rtwo on kinematics decoding in the Maze dataset ($-0.02$ \rtwo against the 6-layer baseline), and a 2-layer NDT matches the 6-layer performance. Note that the shallower NDTs train faster than LFADS again due to parallelism, as parallelism avoids the costly backpropagation through time used to train recurrent networks. In our case, the 6-layer NDT was much larger than the AutoLFADS model; AutoLFADS training times are more appropriately compared with the 2 or 1-layer NDT.

Smaller models may also be more performant in limited data settings. The 2-layer model achieves  0.866 \rtwo when training on just 92 trials (not shown), outperforming the AutoLFADS model. Regularization is still critical for the smaller 2-layer model: performance drops to 0.4 \rtwo when the dropout range is confined to $[0.0, 0.3]$ instead of $[0.2, 0.6]$ (not shown). Though non-exhaustive, this result indicates that the gap between AutoLFADS and NDT when limited to 92 trials (\figref{fig:data_ablate}) may be due to 6-layer models being oversized. Extrapolating beyond this dataset, neural datasets, though smaller than in other domains, may be well-modeled by Transformers so long as the models are appropriately scaled. 

\subsection{Ablative Analysis}
\label{sec:ablative}
\begin{figure}
    \begin{minipage}{0.37\textwidth}
    \centering
      \resizebox{0.9\linewidth}{!}{
    \begin{tabular} {lc}
  \arrayrulecolor{black!20}%
    \toprule
     & $R^2$ $\mathrel{\raisebox{0.1ex}{$(\uparrow)$}}$ \\ 
      \midrule 
      Baseline & $0.918 $\scriptsize{$\pm 0.003$} \\
      \addlinespace[0.1cm]
      \cdashline{1-2}
      \addlinespace[0.1cm]
       Output rates, not logrates & $0.292$ \scriptsize{$\pm 0.104$} \\
       Use “[MASK]” tokens, not zeroes & $0.022$ \scriptsize{$\pm 0.012$} \\
       Constrain dropout $\in [0.0, 0.2]$ & $0.549$ \scriptsize{$\pm 0.053$}
        \\ 
        \bottomrule 
    \end{tabular}
    }
    \captionof{table}{The use of logrates, zero masks, and heavy regularization are all critical to the NDT’s performance. As in main experiments, $\pm$ interval indicates SEM over 3 randomly initialized hyperparameter grid searches.}
    \label{tab:ablative}

    \end{minipage}
    \hspace{0.1cm}
    \begin{minipage}{0.59\textwidth}
    \centering
    \includegraphics[width=0.55\linewidth]{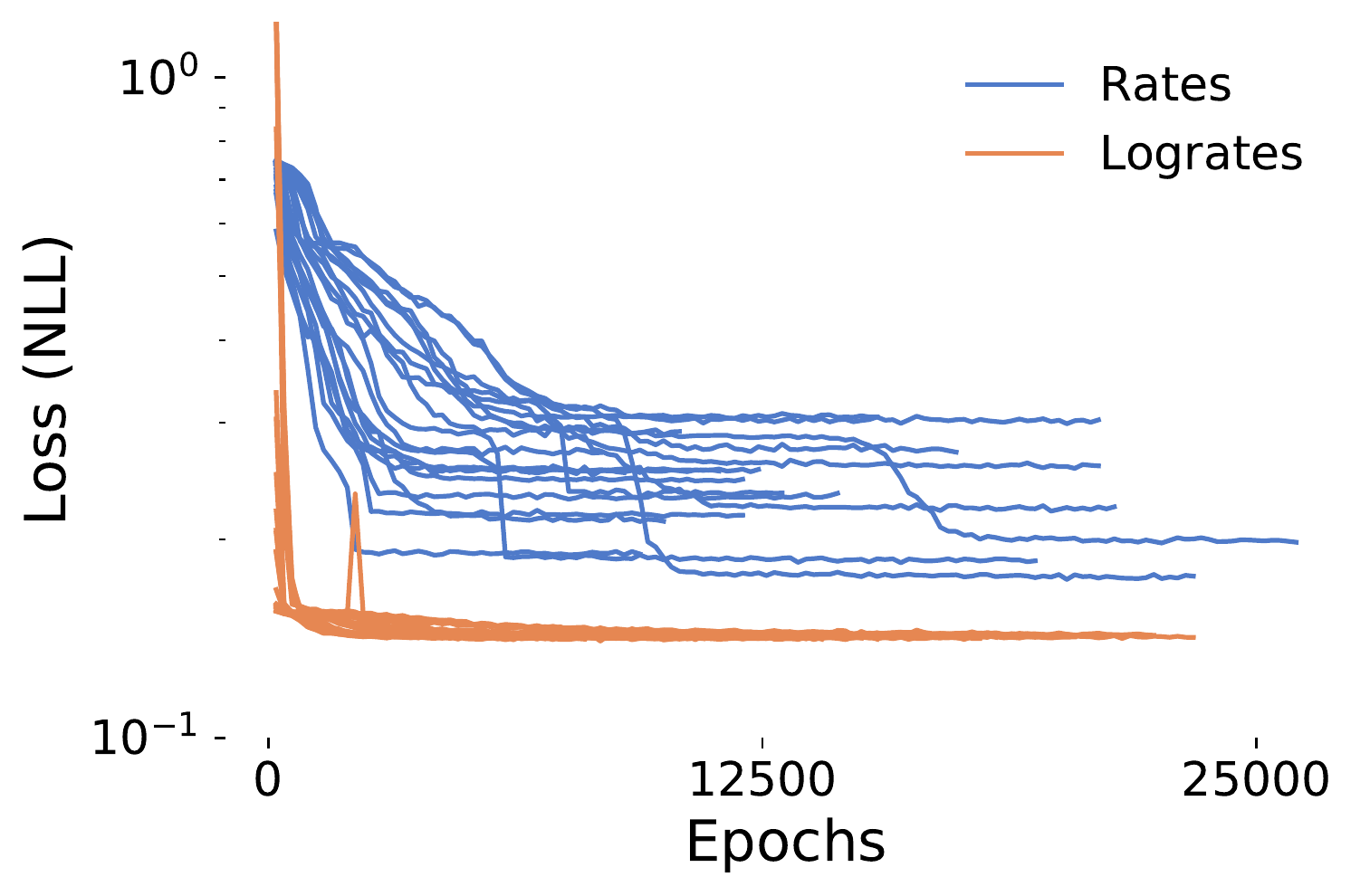}
    \vspace{-0.35cm}
    \caption{\xhdr{Lograte vs. rate inference}. We plot loss trajectories of two hyperparameter sweeps of size 20. One sweep uses firing  (model output is exponentiated before calculating loss) and the other uses firing rates. When using rates, models train less stably and converge to poor solutions.}
    \label{fig:losses}
    \end{minipage}
\end{figure}
We empirically justify three key design choices of the 6-layer NDT by removing them and evaluating the degraded performance on the reaching dataset. Each is critical to achieving high performance (\tabref{tab:ablative}): without these subtle choices, performance is much worse and more variable. For example, models that infer rates instead of logrates train more slowly and fail to converge to a good solution over a wide set of hyperparameters (\figref{fig:losses}).
Notably, inferring rates instead of logrates regresses performance in both 2-layer models and 6-layer models, \ie, constraining models to output in log space is important even with increased capacity. This suggests the NDT cannot learn to exponentiate well, even under long training regimes; this contrasts with findings in~\cite{gao2016linear} where feedforward networks improve on exponential nonlinearities for fitting linear dynamical systems on neural population activity. We also experimented with a few training details relevant in other Transformer works, such as variable length mask spans or adding embedding layers, but found their contributions on the reaching dataset to be marginal, on the order of 1-2\% \rtwo. 
\section{Discussion}
\label{sec:discussion}
We have introduced the NDT, a parallel neural network architecture for neural spiking activity, and shown it can be competitive with RNNs in autonomous dynamical settings while achieving substantially faster inference. Further, with careful architecture choices, the NDT could even match RNN performance on datasets with as few as 92 training trials (0.2 Mb). This indicates that Transformers are compatible with dataset sizes that are typically available in systems neuroscience. \\

\xhdr{The most critical limitation} of the NDT, and thus an important avenue for future work, is its inability to model non-autonomous dynamics, \ie systems with unpredictable external perturbations. This occurs when unmonitored brain areas send signals to the recorded area. 
For example, unpredictable experiment cues that are first processed in somatosensory or visual areas will propagate and perturb the dynamics of recorded motor areas. 
LFADS, which explicitly models inputs for such non-autonomous settings, outperforms NDT by over 20\% \rtwo in a preliminary experiment with a synthetic, non-autonomous dataset, the Chaotic RNN with Inputs studied in~\citet{sussillo2016lfads}. Enabling the NDT to learn sample-efficiently in such non-autonomous settings would likely require architectural modifications to explicitly model both dynamics and inputs.

Despite this limitation, we put forth the NDT as a forward-looking proposal. 
We believe the NDT and more generally the Transformer can benefit neuroscience due to the Transformer's rapid rise in the broader machine learning community. This broader community has advanced and will continue to advance Transformer tooling, analysis, and theory; we provide an overview of recent directions in~\secref{sec:transformer_lit}. Many of these advances could translate to neuroscientific applications. We provide two such examples: 

\begin{itemize}
    \item Story generation requires modeling of both sensible short-term sentence structure and a coherent long-term storyline. While RNNs struggle to learn long-term dependencies, the Transformer's parallel design makes it less biased with respect to either short or long term dependencies. This enables the Transformer to produce long passages of coherent text~\citep{radford2019language}. Analogously, a single Transformer model may yield insights around both fast and slow features of neural activity, uncovering hierarchy within the activity that maps naturally to the multi-scale nature of animal behavior~\citep{berman2016predictability,berman2018measuring}.

    \item Transformers have been productively used to understand the interaction of data from multiple modalities. For example, vision-language transformers~\citep{lu2019vilbert} produce language representations that are contextualized by accompanying images. Similar techniques could be applied to build models which incorporate recordings of multiple brain areas, different recording modalities, and behavioral measurements.

\end{itemize}

However, the major driver of the Transformer's popularity is its ability to scale to large amounts of training data better than RNNs (\ie through faster training). As increasing training data generally improves machine learning models across domains, we anticipate that larger datasets from new recording technologies and dataset aggregation will further improve the NDT’s performance and applicability, possibly past recurrent methods. Notably, these large datasets need not be excessively difficult to collect. For example, they could consist of neural activity that is continuously collected without constrained or even measured behavior. In other domains, the largest datasets tend to be similarly unstructured, naturally-occurring data, such as freeform text extracted from the internet. In a large-scale ``pretraining'' step, networks can learn deep representations of such data in a self-supervised manner, using methods such as those used to train LFADS and NDT. Pretrained representations make subsequent learning for downstream tasks much more data-efficient. The seeming universality of the representations learned in these tasks, for example, has prompted the GLUE language benchmark~\citep{wang2018glue} to assess how well single models perform on 9 different language tasks. An analogous effort in neuroscience may help reveal all the different computational roles of a given neural population, much as prior work has sought to find preferential tuning properties for single neurons.

One promising avenue in the analysis of trained RNNs is the application of techniques from nonlinear dynamical systems theory to interrogate the RNNs’ learned dynamical structure~\citep{sussillo2013opening,sussillo2014neural,golub2018fixed,maheswaranathan2019sentiment}. The Transformer is currently disconnected from these dynamical techniques, as it lacks a recurrent structure to analyze. It would be useful, even beyond the computational neuroscience community, to try to bridge this gap and understand how the Transformer represents dynamical structure.

\section{Acknowledgements}
We thank Andrew Sedler, Yahia Ali, and Ruyi Marone for their insights and conversations. We also thank Krishna Shenoy, Mark Churchland, Matt Kaufman, and Stephen Ryu for sharing the Monkey J Maze dataset. 
The views and conclusions contained herein are those of the authors and should not be interpreted as necessarily representing the official policies or endorsements, either expressed or implied, of the U.S. Government, or any sponsor.  

\section{Author Contributions}
JY and CP jointly contributed towards conceptualization, writing, and revision. JY was responsible for investigation and software, and CP was responsible for funding acquisition and resources.

\section{Methods}
\label{sec:methods}

\subsection{Data Availability}

The Lorenz dataset, along with generation scripts for the Chaotic RNN dataset, are available in the code repo.
The M1 reaching dataset will soon be released at \href{https://gui.dandiarchive.org/\#/dandiset/000070/}{https://gui.dandiarchive.org/\#/dandiset/000070/}~\cite{maze_datarelease}. Our experiments used the recording from 2009-09-18.

\subsection{Architectural Details}
\label{sec:methods_arch}
We provide additional information about the Transformer encoder and self-attention mechanism~\citep{neurips2017vaswani,opennmt2017klein} for more details. Inputs to a Transformer layer pass through a self-attention block, a layer norm block~\citep{ba2016layer}, an MLP, and another layer norm.

\xhdr{Attention Scaling.}
In the original Transformer~\citep{neurips2017vaswani}, the authors found scaling the dot product by a factor $\sqrt{d}$, where $d$ is input dimensionality, improves learning. That is: 
\begin{align*}
    Y &= \text{softmax}(QK^T)V \: \text{becomes} \\
    Y &= \text{softmax}(QK^T/\sqrt{d})V
\end{align*}
Per the implementation of attention available through Pytorch, we maintain this scaling.

\xhdr{Position Embeddings.}
Self-attention lets inputs query for relevant information from other inputs. However, if we directly feed population representations, inputs would be unable to query for information from a particular timestep, \ie there is no intrinsic ordering of the inputs. This is inappropriate in most cases, including ours. To account for input order, we add a learned position embedding (\ie a unique vector representing the identity of the input timestep) to each input before it is fed into the transformer layers.

\xhdr{The rest of the Transformer layer.}
Following self-attention, we have layer normalization and an MLP. Each layer normalization block receives a single population state vector as input and normalizes this input using the mean and variance of its elements. The MLP comprises 2 linear layers joined with a non-linear ReLU activation, and similarly transforms a single input to a single output. Dropout layers are added right before the inputs enter the transformer body (the consecutive transformer layers), right after they exit the body, and right after each linear layer in the MLP of each transformer layer.

\subsection{Further Directions in Transformer Literature.}
\label{sec:transformer_lit}
The Transformer's empirical success has prompted several avenues of further research. One direction, which this work falls under, is the application of the Transformer to new domains, such as for protein folding~\citep{drori2019accurate} or reinforcement learning~\citep{parisotto2019stabilizing}. Other work, often in natural language processing, focuses on improving Transformer performance along some axis. Task metrics can be improved with data engineering and modifying training curricula~\citep{liu2019roberta}, while memory and compute efficiency can be improved with both architectural changes and post-training procedures like pruning or distillation~\citep{tay2020efficient,fournier2021practical}. The Transformer's attention mechanism has also been applied generally in graph neural networks~\citep{Wu2021GNN}. 

A final direction of interest is Transformer interpretability. Transformer models can be analyzed in much the same, possibly contentious, ways as other deep models, through probes of model activations~\cite{tenney2019classical} or through gradient-baseds visualization~\cite{chefer2020transformerInterpretability}. Uniquely, the Transformer is often interpreted with respect to its attention weights. These weights are often intuitive, \eg in language, when an adjective's attention to the preceding contextualizing adverb is high. However, to date there has been no widely accepted theory for how to interpret such attention, and so attention-based analysis is often debated~\cite{wiegreffe2019attention}.

\subsection{Hyperparameters}
\label{sec:methods_hp}
NDT searches are swept over:
\begin{itemize}
    \item Dropout ratio, as described in~\secref{sec:results}.
    \item Context span, the number of timesteps forward and backward each input aggregates information from. Span is swept between 4 and 32 steps in both directions for the synthetic datasets, and 10 and 50 in the reaching datasets.
    \item The ratio of masked tokens that are replaced with a random input instead of a zero mask, and the ratio that are not replaced at all (a methodology from BERT to reduce train-test distribution shift). Zero mask ratio is between $0.5$ and $1.0$ on synthetic datasets, and $0.6$ and $1.0$ on the reaching dataset. Of the remaining masked tokens, between $0.9$ and $1.0$ are replaced with random inputs on synthetic datasets, between $0.6$ and 1.0 on the reaching dataset. 
    \item Length of masked span~\cite{lewis2020bart} is set between 1 and 5 in synthetic datasets, and 1 and 7 in reaching dataset.
\end{itemize}
 
AutoLFADS PBT optimizes over:
\begin{itemize}
    \item Dropout, from $0.0$ to $0.6$
    \item Coordinated Dropout~\citep{neurips2019keshtkaran} rate, from $0.01$ to $0.7$
    \item $L^2$ penalties for the generator from $1\text{e-}4$ to $1.0$
    \item KL penalties for the initial condition from $1\text{e-}5$ to $1\text{e-}3$
\end{itemize}

Both models optimize learning rate, from $1\text{e-}5$ to $5\text{e-}3$. The LFADS controller is kept off as we study autonomous settings. Note that although we find these AutoLFADS settings outperform the ranges reported in~\citep{keshtkaran2021autolfads}, we only claim they are sufficient and not necessary for achieving reported results. PBT settings such as early stopping metrics and epochs per generation are as in~\citep{keshtkaran2021autolfads}. Other hyperparameters are available in the code.

\subsection{Synthetic Dataset}
\label{sec:methods_synth}
The train-val split is 0.8 and 0.2 for each dataset. The Lorenz dataset has 1560 total trials, 50 timesteps, and 29 channels. These trials comprise 65 conditions (firing rate trajectories) with 24 trials sampled per condition. The chaotic RNN dataset is generated with $\gamma=1.5$ and has 1300 total trials (with 100 conditions and 13 trials per condition), 100 timesteps, and 50 channels. \rtwo is calculated by flattening timesteps and trials, and averaging across input channels, as done in~\citep{keshtkaran2021autolfads}.

\subsection{Kinematic Decoding}
\label{sec:methods_kin}
We decode 2D hand velocity from inferred rates at single timesteps using ridge regression with $\alpha=0.01$. As in~\cite{pandarinath2018inferring}, we find improved decoding performance by applying a 90 ms lag between neural activity and the corresponding kinematics, i.e., while rates are inferred between a (-250ms, 450ms) window around movement onset, kinematics are predicted only around (-160ms, 450ms). 

\subsection{Timing Tests}
\label{sec:methods_time}
We report the time of a forward pass through the NDT and LFADS models, i.e. the time it takes to infer rates from spike inputs. 1 posterior sample is used for LFADS. Times are averaged over 1300 trials. Measurements were taken on a machine (on CPU) with 32GB RAM and a 4-core i7-4790K processor running at 4.2 GHz.

{\small
\setlength{\bibsep}{0pt}
\bibliography{bib/strings,bib/main}
}

\end{document}